\documentclass{sigchi}

\CopyrightYear{2020}
\setcopyright{acmlicensed}
\doi{http://dx.doi.org/10.1145/3357236.3395515}
\isbn{978-1-4503-6974-9/20/07}
\conferenceinfo{DIS 20}{July 6--10, 2020, Eindhoven, Netherlands}
\acmPrice{15.00}

\toappear{\scriptsize Permission to make digital or hard copies of all or part of this work for personal or classroom use is granted without fee provided that copies are not made or distributed for profit or commercial advantage and that copies bear this notice and the full citation on the first page. Copyrights for components of this work owned by others than the author(s) must be honored. Abstracting with credit is permitted. To copy otherwise, or republish, to post on servers or to redistribute to lists, requires prior specific permission and/or a fee. Request permissions from permissions@acm.org. \\

This is the pre-print version. The paper is to appear in the proceedings of the DIS 2020 conference. Final version DOI: https://doi.org/10.1145/3357236.3395515.}

\usepackage{balance}       
\usepackage{graphics}      
\usepackage[T1]{fontenc}   
\usepackage{txfonts}
\usepackage{mathptmx}
\usepackage[pdflang={en-US},pdftex]{hyperref}
\usepackage{color}
\usepackage{booktabs}
\usepackage{textcomp}
\usepackage{braille}

\usepackage{microtype}        
\usepackage{ccicons}          

\def\plaintitle{HaptiRead: Reading Braille as Mid-Air Haptic Information}

\def\emptyauthor{}
\def\plainkeywords{Mid-air Haptics, Ultrasound, Haptic Feedback, Public Displays, Braille, Reading by Blind People.}

\makeatletter
\def\url@leostyle{%
  \@ifundefined{selectfont}{
    \def\UrlFont{\sf}
  }{
    \def\UrlFont{\small\bf\ttfamily}
  }}
\makeatother
\def\pprw{8.5in}
\def\pprh{11in}

\definecolor{linkColor}{RGB}{6,125,233}
\hypersetup{%
  pdftitle={\plaintitle},
  pdfauthor={\emptyauthor},
  pdfkeywords={\plainkeywords},
  pdfdisplaydoctitle=true, 
  bookmarksnumbered,
  pdfstartview={FitH},
  colorlinks,
  citecolor=black,
  filecolor=black,
  linkcolor=black,
  urlcolor=linkColor,
  breaklinks=true,
  hypertexnames=false
}

\begin{document}

\title{\plaintitle}

\numberofauthors{1}
\author{%
  \alignauthor{Viktorija Paneva \hspace{2cm} Sofia Seinfeld \hspace{2cm} Michael Kraiczi	\hspace{2cm} J{\"o}rg M{\"u}ller\\
\affaddr{University of Bayreuth, Germany\\
\{viktorija.paneva, sofia.seinfeld, michael.kraiczi, joerg.mueller\}@uni-bayreuth.de}}
}

\teaser{ 
\centering 
\includegraphics[width=\textwidth, height=5cm]{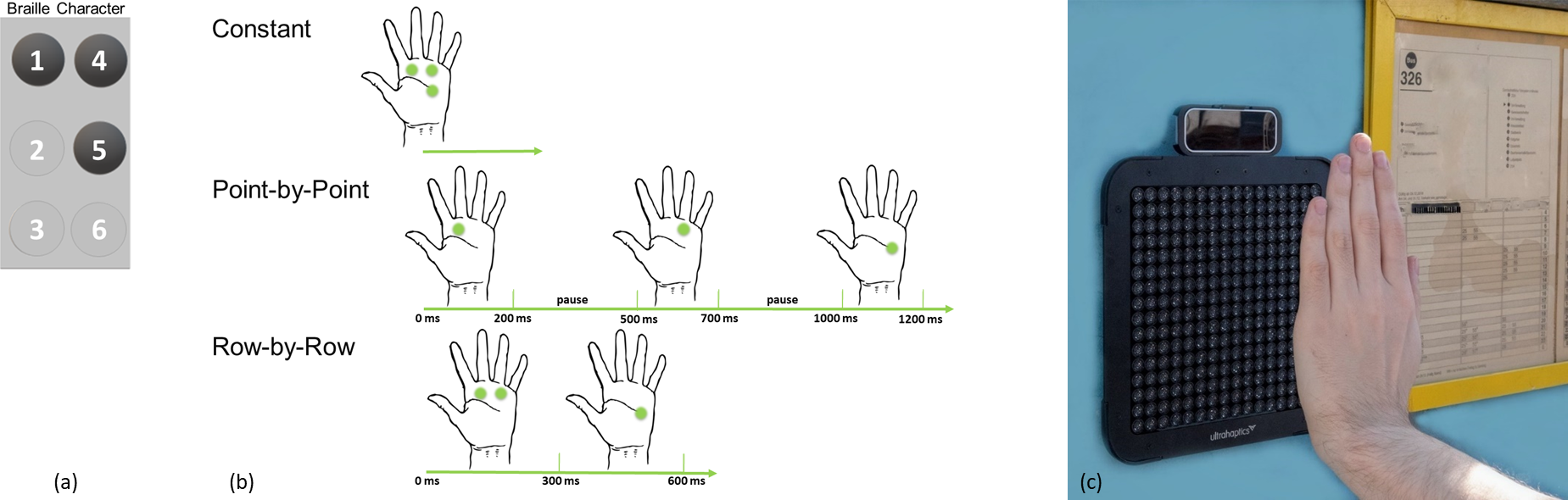} 
\caption{With HaptiRead we evaluate for the first time the possibility of presenting Braille information as touchless haptic stimulation using ultrasonic mid-air haptic technology. We present three different methods of generating the haptic stimulation: Constant, Point-by-Point and Row-by-Row. (a) depicts the standard ordering of cells in a Braille character, and (b) shows how the character in (a) is displayed by the three proposed methods. HaptiRead delivers the information directly to the user, through their palm, in an unobtrusive manner. Thus the haptic display is
 particularly suitable for messages communicated in public, e.g. reading the departure time of the next bus at the bus stop (c).} 
\label{fig:teaser} } 

\maketitle


\begin{abstract}
  Mid-air haptic interfaces have several advantages - the haptic information is delivered directly to the user, in a manner that is unobtrusive to the immediate environment.
  They operate at a distance, thus easier to discover; they are more hygienic and allow interaction in 3D. 
We validate, for the first time, in a preliminary study with sighted and a user study with blind participants, the use of mid-air haptics for conveying Braille. 
We tested three haptic stimulation methods, where the haptic feedback was either: a) aligned temporally, with haptic stimulation points presented simultaneously (Constant); b) not aligned temporally, presenting each point independently (Point-By-Point); or c) a combination of the previous methodologies, where feedback was presented Row-by-Row. 
The results show that mid-air haptics is a viable technology for presenting Braille characters, and the highest average accuracy ($94\%$ in the preliminary and $88\%$ in the user study) was achieved with the Point-by-Point method.
\end{abstract}




\section{Introduction}

There are several challenges that blind people face when engaging with interactive systems in public spaces.
Firstly, it is more difficult for the blind to maintain their personal privacy when engaging with public displays.
Audio feedback is easily overheard by bystanders and can be perceived as obtrusive, since it contributes to the environmental noisescape.
Some interfaces, such as ATMs, feature a headphone plug.
In this case, however, users need to remember to bring headphones and once they start the interaction, they might have more difficulty monitoring events in their surroundings.
Refreshable Braille displays, consisting of lines of actuated pins, also have some shortcomings.
The information they can convey is limited to patterns of dots, which is suitable for text, but not sufficient for content involving shapes and objects (e.g. data charts). 
It can be difficult to detect them from a distance, since the user has to already touch them to know they are there.
The physical contact with these interfaces could potentially cause hygiene problems in public spaces, e.g. hospitals.
They contain moving parts, which can become clogged by dirt in public spaces.

As a potential solution for these challenges, we present HaptiRead - a concept of the first public display that presents Braille information as touchless stimulation, using mid-air haptic technology \cite{ultrahaptics}.
The feedback generated by HaptiRead is delivered directly to the user, without disturbing the environment. 
The system can detect the user's hand using a built-in Leap Motion sensor, and render the Braille text where the hand is, improving detectability.
Its contactless nature means that it could prevent hygiene-related issues.
Because it contains no moving parts, it is potentially more robust for public spaces. 
The combination of easily-detectable, yet unobtrusive interface could potentially encourage more blind people to use the accessibility feature, thus granting them more autonomy and independence in their daily life. 
Lastly, the volumetric interaction space of the interface, allows for content versatility, beyond Braille text.

To our knowledge, there are no formal studies exploring the potential of using mid-air haptic technology to convey Braille information.
In this paper, we evaluate different methods for presenting the haptic stimuli in mid-air, in an iterative design process.
Then we test with users the three most promising methods: Constant (emission of all haptic points at the same time), Point-by-Point, and Row-by-Row.
We first conduct a preliminary study with sighted participants, investigating whether HaptiRead can provide enough haptic cues to differentiate between different dot patterns.
Then we evaluate the performance and user experience in a user study with blind participants, proficient in Braille. 

Our main contributions are:

1. We present the first user study that investigates the use of a mid-air haptic interface with blind participants.

2. Through user studies, we demonstrate that it is possible to effectively distinguish between different Braille patterns, where each dot of the pattern is represented by a mid-air haptic stimulation point.

3. We present and compare three different haptic stimulation methods for generating Braille characters in mid-air.

\section{Related Work}

\subsection{Mid-air Haptics}
Ultrasonic mid-air haptics~\cite{iwamoto, ultrahaptics} is a technology that allows for haptic feedback to be projected directly onto users' unadorned hands.
Focused ultrasonic waves are emitted by a phased array of transducers at a frequency of $40$~kHz.
By modulating the waves with a frequency detectable by the receptors in the human skin, it is possible to create a perceivable haptic sensation in mid-air~\cite{ultrahaptics}. 

Early prototypes were able to generate a single haptic point using linear focusing~\cite{hoshi}.
Wilson et al.~\cite{wilson} investigated the perception of an ultrasonic haptic point experimentally. 
Alexander et al.~\cite{alexander} introduced multi-point haptic feedback, using spatial and temporal multiplexing. 
Later prototypes used optimization algorithms to generate multiple haptic points~\cite{gavrilov,ultrahaptics}. 
User studies involving multi-point haptic feedback, carried out by Carter et al.~\cite{ultrahaptics}, show that differentiability between two haptic points improves, when they are modulated with different frequencies, and the accuracy of determining the correct number of points increases with the distance - for distances of $3$~cm and above the accuracy was over $85\%$. 
An algorithm for creating volumetric shapes using the mid-air haptic technology was presented by Long et al.~\cite{long}. 
Haptogram~\cite{korres} is an alternative method to generate 2D and 3D tactile shapes in mid-air, using a point-cloud representation.
In addition to points and shapes, a rendering technique for the creation of haptic textured surfaces has been demonstrated in~\cite{freeman}. 

By tuning parameters, such as the location of the mid-air haptic stimulus, number of haptic points, modulating the frequency, among other factors, it is possible to generate haptic patterns that are suitable for different applications.
For example, Vi et al.~\cite{vi} used  mid-air haptic technology to enhance the experience of visual art in a museum, Martinez et al.~\cite{martinez} generated haptic sensations that mimic supernatural experiences in VR and in~\cite{harrington} buttons and sliders were augmented with mid-air haptic feedback in a driving simulator, to reduce off-road glance time.
Gil et al.~\cite{whiskers} explored the perception of mid-air haptic cues on the face, across different parameters and in a practical notification task.

\subsection{Braille Interfaces}
Today Braille is mostly read from a nonrefreshable embossed medium (e.g. paper~\cite{braille}). 
Refreshable Braille displays, made of actuated plastic or metal pins, embody a more flexible, but also a more pricey alternative, ranging up to $10 000\$$\footnote{https://canasstech.com/collections/blindness-products/braille-displays}.  

In the past, several methods have been developed for reading Braille on mainstream devices, such as mobile phones and tablets. 
Rantala et al.~\cite{Rantala} presented three interaction methods: scan, sweep and rhythm for reading Braille on mobile devices with a touchscreen. 
Al-Quidah et al.~\cite{morsebraille} optimized the temporal rhythm method further, by developing an encoding scheme for each possible column combination in a Braille character, similar to the Morse code.
The encoding scheme lowers the time it takes to represent a character. 
The users are required, however, to learn a new mapping. 
The accuracy ranged from $61$ to $73\%$.
\emph{HoliBraille}~\cite{holibraille} is a system consisting of six vibrotactile motors and dampening elements that can be attached to mobile devices in order to enable interaction in Braille, in the form of multipoint localized feedback.
Another method for presenting Braille characters on a mobile phone is \emph{VBraille}~\cite{vbraille}. 
The touchscreen of the phone is divided into six cells in the usual Braille order (Figure~\ref{fig:teaser}(a)).
When a cell representing a raised dot is touched, the phone vibrates. 

\emph{UbiBraille}~\cite{ubibraille} is a wearable device consisting of six aluminum rings that transmit vibrotactile feedback. 
The device is able to simultaneously actuate the index, middle and ring finger of both hands of the user, each corresponding to one Braille cell. 
Luzhnica et al.~\cite{luzhnica} investigated encoding text using a wearable haptic display, in a hand, forearm and two-arms configuration.
Tactile information transfer on the ear was explored with \emph{ActivEarring}~\cite{ActivEarring}, a device able to stimulate six different locations on the ear using vibration motors.

\subsection{Summary}
In the past, as an alternative to Braille displays consisting of individually actuated pins, a variety of methods and devices relaying on vibrotactile feedback have been researched.
The area of touchless mid-air haptics for Braille applications has been unexplored up to now.
In this paper we propose HaptiRead, an interface for blind users with the potential to provide improved privacy, detectability, hygiene and variability of displayable content.

\section{The System}
For providing the haptic feedback in mid-air we use the Stratos Explore development kit from Ultraleap\footnote{www.ultraleap.com}. 
The hardware is equipped with 256 transducers, that emit ultrasonic waves to create up to eight perceivable points at a maximum range of approx. $70$~cm, as well as a Leap Motion hand tracking module. 
The board's update rate for the ultrasound is $40$~kHz, which implies that the diameter of the generated points is $8.6$~mm (the wavelength of sound at $40$~kHz). 
Such high frequencies are above the threshold of human tactile perception in the hand~\cite{dalecki}.
Thus the ultrasonic waves are modulated, using frequencies between $100$ and $200$~Hz (recommended by the manufacturer).
For better differentiability, we modulate each haptic focus point, representing a different cell in a Braille character, with a different frequency.
We chose a modulation frequency of $200$~Hz for cell 1, $140$~Hz for cell 2, $120$~Hz for cell 3,$160$~Hz for cell 4,$180$~Hz for cell 5 and $100$~Hz for cell 6 (see Figure~\ref{fig:teaser}(a) for ordering convention). 
For consistency, the chosen modulation frequency for each cell was fixed throughout all characters.
For our application we chose a distance of $3$~cm between the centers of the points, since in our pilot tests, it showed the best trade off between the overall size of the pattern and the ability to detect single points. 
 
\section{Iterative Design Process}

\subsection{Interview with a Braille Teacher} 
To gain expert feedback on the HaptiRead concept, we conducted an exploratory interview with a local Braille teacher, with $20$ years of teaching experience.
In her daily life, the teacher uses a mixture of various voice systems, as well as refreshable Braille lines and books.
She relies on Braille for completing tasks that require precision, like reading a phone number or correcting a text.
She responded favourably to the HaptiRead concept and appreciated the compactness and mobility of the device. 
The teacher could envision using it in public spaces for reading timetables, menus in a restaurant or doctor's prescriptions. 
In all of these cases, text-to-speech devices are not suitable, because they can be overheard by bystanders and other existing solutions require the user to have a minimum amount of visual capability.
The teacher suggested that the HaptiRead device might be especially useful for young, congenitally blind people, for training the recognition of dots and their location,  in the Braille learning process.

\subsection{Pilot Study}
We carried out a brainstorming session where different methods, specially designed for presenting Braille characters on a mid-air haptic device, were generated.
The refined list of potential methods to display Braille via touchless haptics is presented in Table~\ref{methods_table}. 
These methods were evaluated in a pilot study with six sighted participants ($1$ female, $5$ male) with no previous experience in Braille or with mid-air haptic systems. 
Most of them reported they felt more comfortable using the methods where part of the pattern or the individual points are sequentially presented. 
For these methods, they reported higher levels of confidence in their ability to correctly identify the patterns. 
The preferred methods were Row-by-Row and Point-by-Point. 
In iterative tests with other pilot participants, we determined the best timespans for displaying the feedback for these methods. 
In the Point-by-Point method, the best results were achieved when individual dots were displayed for $200$~ms, with a $300$~ms pause between subsequent dots and a $500$~ms pause at the end of a character.
Performance with the Row-by-Row method was the best, when the rows were displayed in $300$~ms intervals.
An illustration of the pattern presentation timelines per method, is given in Figure~\ref{fig:teaser}(b).

\subsection{Interview with a Proficient Braille Reader} 
When presented with the haptic stimulation methods in Table~\ref{methods_table}, the interviewee reported that the method rendering all the haptic points simultaneously (Constant), was the most in line with her expectations of reading Braille.  
In addition, the process of transferring her previous Braille knowledge onto the novel system was the most fluent with this method. 
She also stated, however, that her fluency improved rapidly (with the other methods as well)  after a few training sessions.
In her opinion, the interface could particularly be useful for Braille beginners, for whom the refreshable Braille lines are too fast. 
With HaptiRead they can take the time to explore the individual dots and patterns.
In a later consultation, the Braille teacher also stated the Constant method as her preferred one.

\begin{table}[h]
\centering
\begin{tabular}{|l|l|}
\cline{1-2}
\textbf{Method} & \textbf{Description} \\ \cline{1-2} \cline{1-2} \cline{1-2}
Constant & all dots are simultaneously displayed\\ \cline{1-2}
Pulsating & the dots are flashing in sync\\ \cline{1-2}
Rotating & each dot is rotating clockwise\\ \cline{1-2}
Expanding & the dots move away from each other\\ \cline{1-2}
Varying Intensity & dot intensity is fluctuating over time\\ \cline{1-2}
Row-by-Row & rows are subsequently displayed\\ \cline{1-2}
Column-by-Column & columns are subsequently displayed\\ \cline{1-2}
Point-by-Point & only one dot is displayed at a time\\ \cline{1-2}
Morse-Like & dots are presented in a time\\
& sequence, at the same position\\  \cline{1-2}
\end{tabular}
\caption{Haptic stimulation methods evaluated during the design phase.}
\label{methods_table}
\end{table}

\section{Pre-Study}
\begin{figure}[h]
\centering 
\includegraphics[width=\columnwidth,]{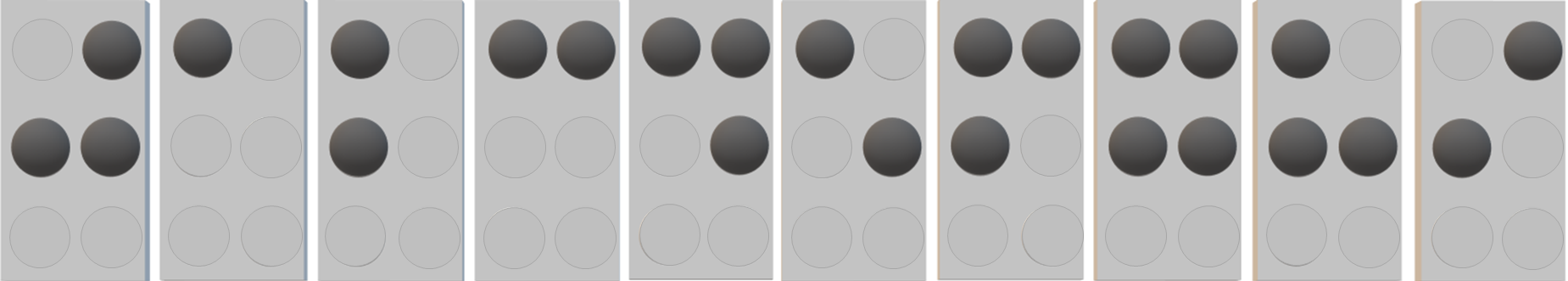} 
\caption{Visual representation of the dot patterns tested. } 
\label{fig:patterns}
\end{figure}

We conduct a pre-study with eighteen sighted participants (11 females and 7 males; 2 left and 16 right-handed), aged between 20 and 40 years (mean 25.29, SD 5.12), to test whether the HaptiRead system provides enough haptic cues for dot pattern recognition.
The participants reported no previous experience with mid-air haptics or knowledge in Braille.
The experimental task was chosen after careful consideration and consultation with Braille experts. 
As this was the first time the participants came in contact with mid-air haptic technology, in order to avoid overwhelming the user with the study protocol, we opted for a simple experimental task that ensures high internal validity and experimental control. 
The task consisted of correctly identifying a pattern of dots being presented in the form of mid-air haptic stimulation.
The possible patterns  were limited to 4-cell Braille characters (see Figure~\ref{fig:patterns}).
Using the methods identified as most promising, in the pilot study - Row-by-Row and Point-by-Point, and in the interview - Constant, the participants were presented with ten dot patterns per method ($30$ trials in total).

To avoid potential learning and ordering effects, a fully counterbalanced design was used.
Each participant was seated in front of a computer and asked to place their left hand $20$~cm above the ultrasonic array and focus on perceiving the pattern presented on their palm. 
They were provided with earmuffs to prevent auditory influences.
Participants had to indicate which pattern they were perceiving on their left hand by selecting the visual equivalent, i.e. visual representation of the pattern, on a screen.
Before completing the actual trials of the study, participants underwent a training session that included four trials for each haptic stimulation method. 
No time limit was given on the response time for each trial, however the time taken to answer was recorded for each trial. 
In the training trials, performance feedback was given. 
The actual trials of the study did not include feedback, so participants were not aware of their performance. 
The experiment lasted approximately $30$~min in total per participant.
The pre-study was approved by the Ethical Committee of University of Bayreuth
and all participants received monetary compensation for their participation.


\subsection{Results}
The average pattern recognition accuracy rate over all methods was $86$\%. 
The highest average accuracy score of $94$\% (SD $7$) was achieved with the Point-by-Point method, whereas for both the Constant and the Row-by-Row method the average score was $82$\% (SD $19.21$).
The average time it took the participants to recognize a pattern was  $10.89$~s (SD $4.75$) for the Constant, $8.55$~s (SD $2.36$) for the Point-by-Point, and  $10.55$~s (SD $4.34$) for the Row-by-Row method.
Note that the participants were instructed to focus on correctly identifying the dot patterns, rather than providing fast answers. 


%
%

The high accuracy rates indicate that it is possible to communicate different dot patterns as touchless haptic stimulation, using all three methods. 
Using the Friedman test, no significant difference was found between the haptic stimulation methods ($\chi^2=5.15$, $df=2$,  $p=0.059$) and in the Mean Time to Respond ($\chi^2=3.11$, $df=2$,  $p=0.21$).
 
\begin{figure}[h]
\centering 
\includegraphics[width=0.6\columnwidth,]{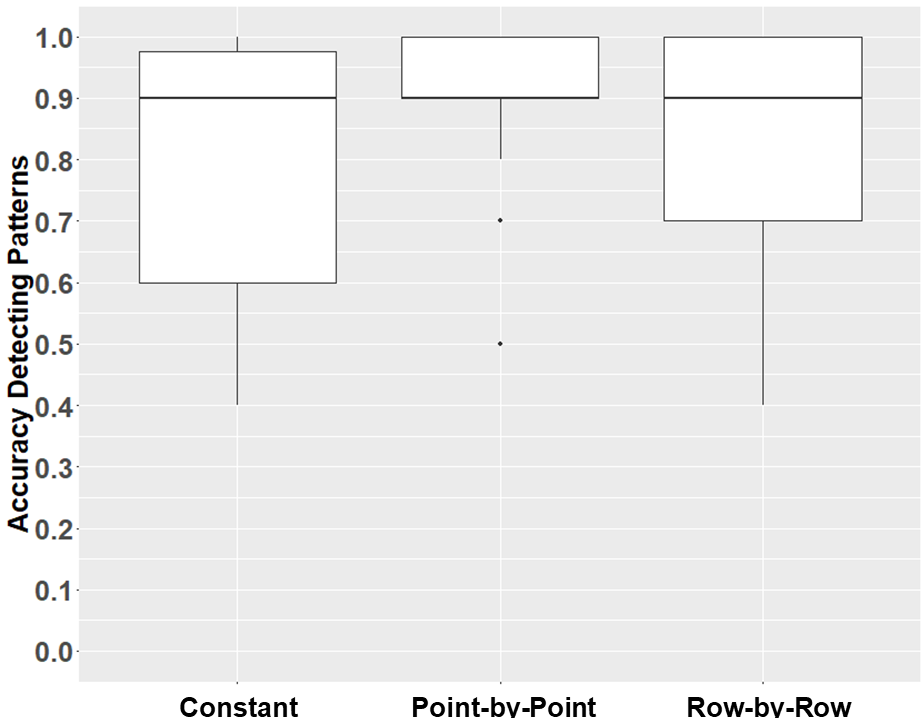} 
\caption{Boxplot of the Accuracy for the three haptic stimulation methods (Constant, Point-by-Point and Row-by-Row) in the pre-study.} 
\label{fig:accuracy_mean_time}
\end{figure}

\section{User Study}
Since in the pre-study all three haptic stimulation methods showed potential to be used for dot pattern presentation, we test all of them in a user study with blind participants. 

\subsection{Experimental Design}
The user study consisted of a within-groups experimental design. 
The participants experienced three possible types of haptic stimulation: 1) Point-by-Point, 2) Constant, and 3) Row-by-Row.
The different methods were presented in a randomized order.

 \begin{table}[b]
 \resizebox{\columnwidth}{!}{
 \begin{tabular}{l|c|c|c|c|c|c|c|c|c|c|c}
ID & 1 & 2 & 3 & 4 & 5 & 6 & 7 & 8 & 9 & 10 & 11 \\
Gender & m & f & f & m & f & m & f & m & m & f & m \\
Handedness & r & r & r & l & r & l & r & r & r & l & r \\
Age & 19 & 45 & 36 & 46 & 52 & 48 & 45 & 29 & 31 & 41 & 70 \\
BE in Years & 13 & 34 & 30 & 28 & 2 & 41 & 40 & 24 & 22 & 35 & 56 \\

 \end{tabular}}
 \caption{Demographic data of the participants. BE = Braille Experience}
 \label{demographics}
 \end{table}

\subsection{Participants}
Eleven blind participants (5 females and 6 males) aged between 19 and 70 (mean 42, SD 13.45) were recruited for the experiment. 
Their demographic data is given in Table~\ref{demographics}.
Before the experiment started, the participants were read basic information about the study and they signed a consent form.
The study was approved by the Ethical Committee of University of Bayreuth 
and followed ethical standards as per the Helsinki Declaration. 
All participants received a monetary reimbursement for their participation.
 
\subsection{Measures and Procedure}
To better accommodate participants' needs, the study was conducted in the familiar environment of their homes. 
All potential distractions (e.g. phones) were removed from the vicinity. 
The participant was comfortably seated and the HaptiRead interface was placed on a table in front of them. 
The participant was encouraged to raise any questions regarding the study and the technology.
After the consent form was signed, a demographic questionnaire was verbally administered. 
Then the participant was asked to complete a short task to verify their proficiency in reading Braille.
The task consisted of five 5-digit numbers in Braille, that they had to read out loud. 
Next, the participant was instructed to place their dominant hand $20$~cm above the ultrasonic array and focus on perceiving the haptic sensation on the palm of their hand. 
The participant was asked to wear headphones during the experiment, to control for any potential auditory influence on their responses. 
Similarly as in the pre-study, before completing the actual trials of the study, the participant underwent a training session that included four trials for each haptic stimulation method. 
The experimental task consisted of a random presentation of trials (10 trials per method, 30 in total), where the participant had to identify the Braille digit presented via mid-air haptics. 
No time limit to respond was given, however the time taken to answer was recorded for each trial. 
The participant was permitted to actively explore the haptic sensation, but instructed to approximately keep the recommended vertical distance to the array.
When the participant recognized the Braille pattern, they stated the corresponding character out loud. 
At this moment the timer was halted, but the feedback continued. 
After completing the experiment, the participant was asked to indicate their subjective opinion of how mentally demanding the task was, as well as how comfortable they felt using each of the haptic stimulation methods.
The questions were answered on a 7 point Likert scale (1 meaning \textit{not mentally demanding at all/not comfortable at all}, 7 meaning \textit{extremely mentally demanding/extremely comfortable}).  
Next, the System Usability Scale~\cite{brooke} questionnaire was verbally administered. 
The participant was asked to answer the questionnaire considering the HaptiRead system with their preferred haptic stimulation method.
Finally, a semi-structured interview was conducted.
The user study lasted approximately one hour per participant.

\subsection{Results}

\subsubsection{Accuracy and Time to Respond}
\begin{figure}[h]
\centering 
\includegraphics[width=0.6\columnwidth,]{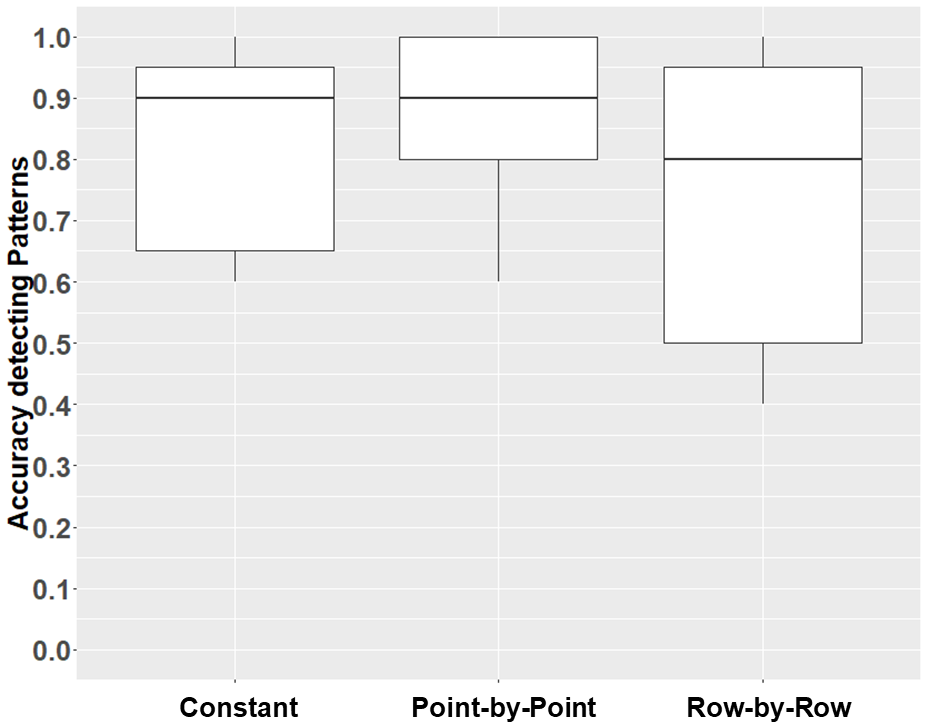} 
\caption{Boxplots of the Accuracy for the three haptic stimulation methods (Constant, Point-by-Point and Row-by-Row).} 
\label{fig:accuracy_mean_time_blind}
\end{figure}

The average accuracy was  $81$\% (SD $17$) for the Constant, $88$\% (SD $14$) for the Point-by-Point, and  $75$\% (SD $23$) for the Row-by-Row method.
Figure~\ref{fig:accuracy_mean_time_blind} shows that the Point-by-Point method achieved the highest mean accuracy score, followed by the Constant, and the Row-by-Row method. 
Using the Friedman test, no significant difference in the Accuracy between the haptic stimulation methods was found ($\chi^2=4.92$, $df=2$,  $p=0.08$). 
The mean time to identify a character totals $7.19$~s (SD $4.02$) for the Constant, $7.30$~s (SD $2.44$) for the Point-by-Point and $7.31$~s (SD $3.45$) for the Row-by-Row method.
The Friedman test indicated no significant differences between the three ($\chi^2=1.64$, $df=2$, $p=0.44$).

 
\subsubsection{Mental Demand and Perceived Comfort}
\begin{figure}[h]
\centering 
\includegraphics[width=\columnwidth,]{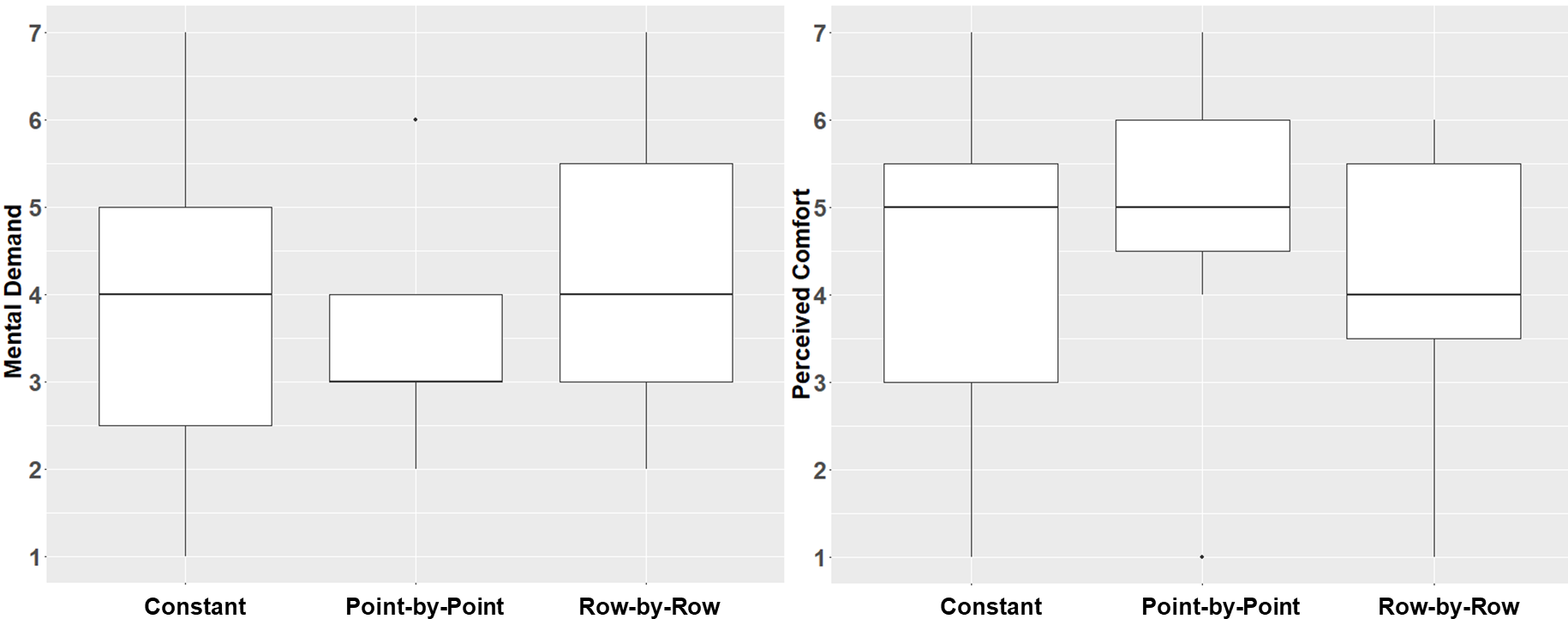} 
\caption{Boxplot of Mental Demand and Perceived Comfort for the three haptic stimulation methods (Constant, Point-by-Point and Row-by-Row).} 
\label{fig:mental_demand_comfort}
\end{figure}

On average, the participants reported slightly lower Mental Demand when using the Point-by-Point method (median $ = 3$) to read the Braille characters, compared to the Constant and Row-by-Row methods (median $ = 4$ for both).
Lower levels of Comfort were reported for the Row-by-Row method (median $ = 4$), compared to the Constant and Point-by-Point method (median $ = 5$ for both).
The scores are presented in Figure~\ref{fig:mental_demand_comfort}.
However, using the Friedman test, no significant difference for Mental Demand ($\chi^2=2.34$, $df=2$, $p=0.30$) or Perceived Comfort ($\chi^2=1.90$, $df=2$, $p=0.39$) was found.

\subsubsection{Confusion Matrix Analysis}
\begin{figure}[h]
\centering 
\includegraphics[width=0.75\columnwidth,]{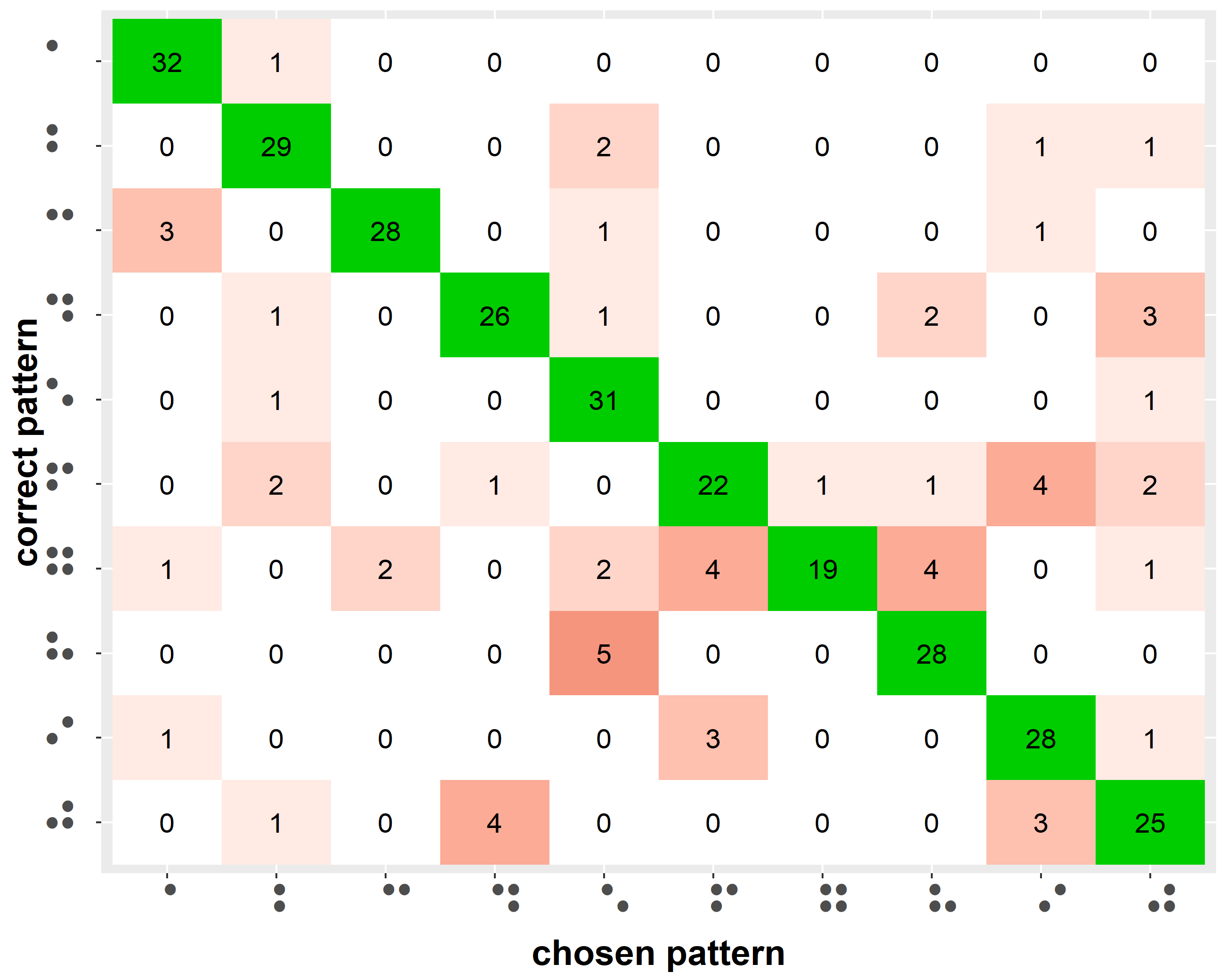} 
\caption{Visualization of the correctly and incorrectly identified patterns in the user study with the respective frequencies.} 
\label{fig:confusion_matrix}
\end{figure}

The confusion matrix, providing information about the most frequently mistaken patterns, is shown in Figure~\ref{fig:confusion_matrix}. 
The pattern \braille{7} was identified correctly the least amount of times ($19$ out of $33$), whereas the pattern \braille{1} consisting of only one haptic point was identified correctly almost always ($32$ out of $33$ trials).
The pattern \braille{7} was most often mistaken for \braille{6} and \braille{8}.
The majority of the errors ($61\%$), occurred due to misperception of a single haptic stimulation point.
In 30 trials, the error was due to a false negative (e.g. \braille{8} identified as \braille{5}), and in 8 trials, due to a false positive (e.g. \braille{9} identified as \braille{6}).
In $31\%$ of the errors, both a false positive and false negative occurred (e.g. \braille{4} identified as \braille{0}).  
The remaining  $8\%$ of the errors, were due to the omission of two or more points, i.e identifying \braille{7} as \braille{1} and \braille{3}.
The confusion matrices per method, in the pre-study and the user studies, are provided in the supplementary material.

\subsubsection{System Usability}
\begin{figure}[h]
\centering 
\includegraphics[width=0.8\columnwidth,]{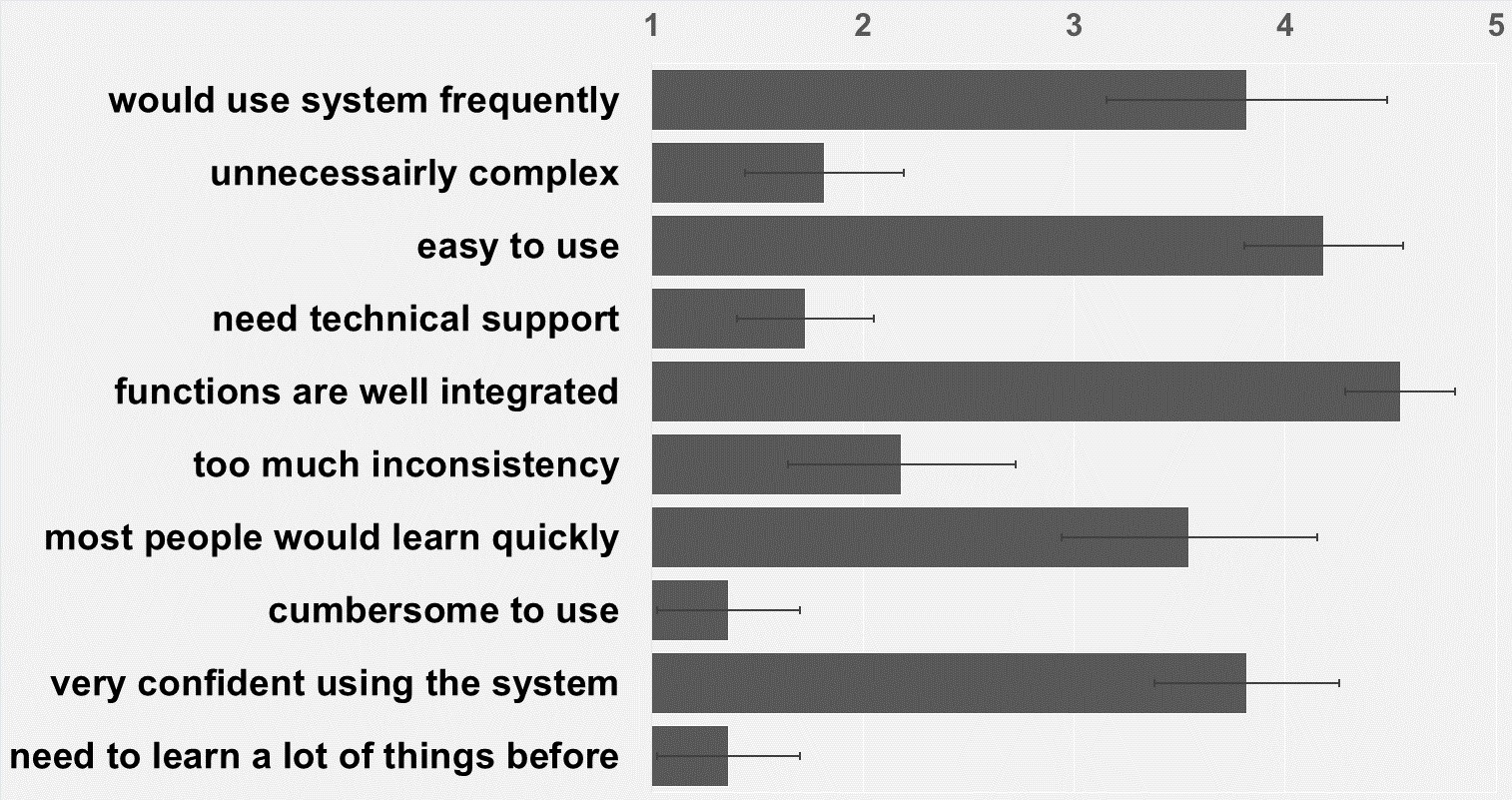} 
\caption{Scores on the System Usability Scale for the HapiRead system (1 = strongly disagree, 5 = strongly agree, n = 11).} 
\label{fig:SUS}
\end{figure}

The results of the System Usability Questionnaire are shown in Figure~\ref{fig:SUS}. 
The HaptiRead system scored a SUS score of $78.6$ (SD $7.6$), meaning that the participants rated the system as above average in terms of usability. 
The main concern participants expressed was that, for example, older members of the blind community might not be able to learn quickly how to use and operate the interface.

\subsubsection{Semi-Structured Interview}
There was no overall preferred haptic stimulation method by the participants in the user study. 
Some liked the familiarity of the Constant method: \textit{the Constant method felt closest to regular Braille}.
Others felt more comfortable with the temporally modulated methods because they provided them with additional spatial cues: \textit{I liked that with the Row-by-Row method I had an indication how to orient my hand.} 
Regarding the Point-by-Point method, participants stated: \textit{It was more like a flow and the image was constructed over time, like an image is constructed over time in real life.}; \textit{I found the method too slow, but I felt the most secure}.
Most participants reported they felt comfortable with using the palm, instead of the finger, for reading the Braille numbers.
None of the participants reported feelings of fatigue. 
As scenarios where they would use HaptiRead, participants listed: to read door signs in public spaces, at the self-checkout register in the supermarket, at a ticket machine, as a small portable clock, to read relief maps etc.
One participant stated: \textit{I could use it at work for the punch clock, to see the time I worked}, and another: \textit{I could imagine using the system at home, because my mechanical Braille lines got too slow over time.}

\section{Discussion}

\subsection{Reading Braille with Mid-Air Haptics}
\begin{figure}[t]
\centering 
\includegraphics[width=0.8\columnwidth,]{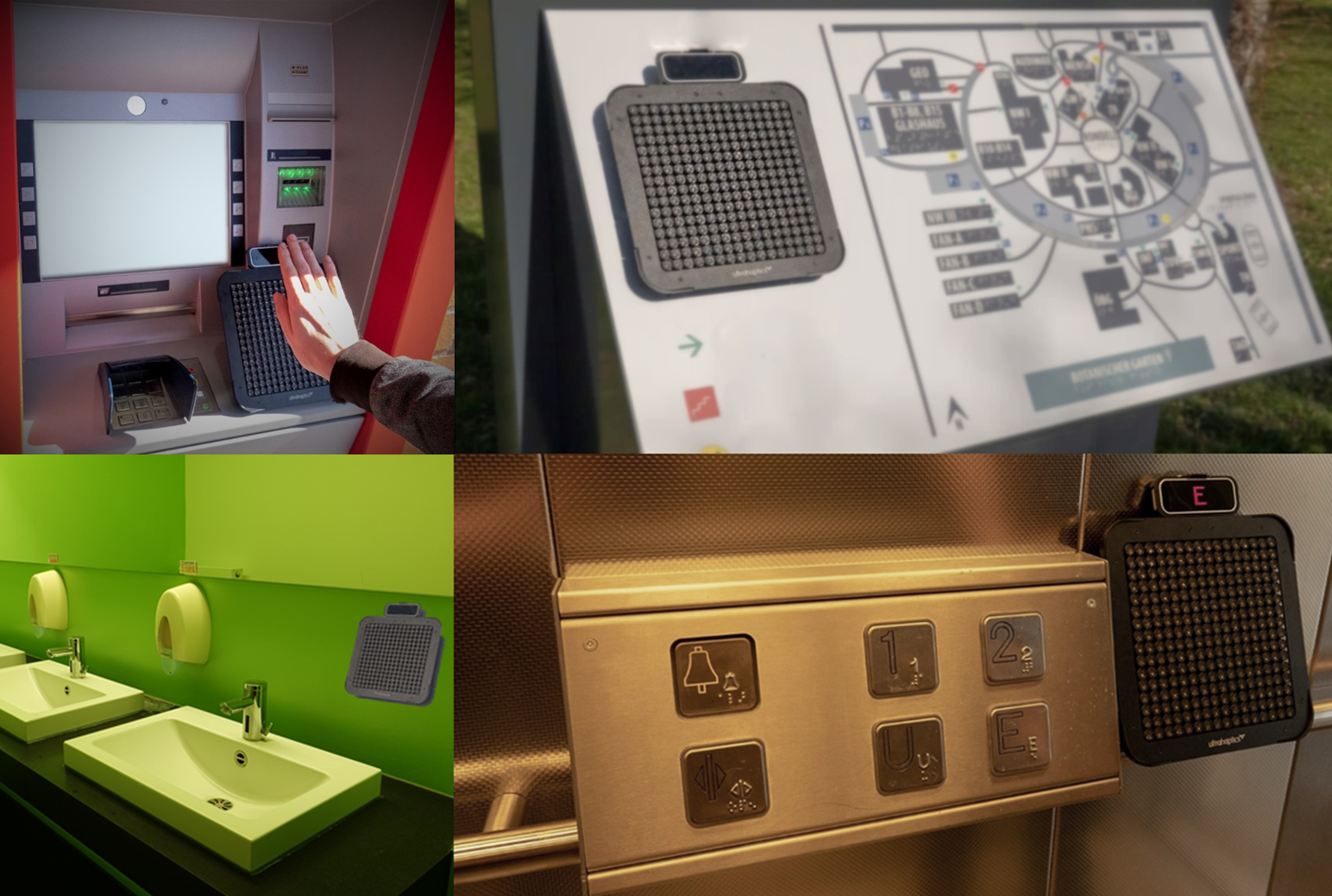} 
\caption{Potential applications scenarios for HaptiRead (left to right, top to bottom): to read the account balance at the ATM, display landmark names and direction on navigation maps, to facilitate item localization in restrooms, and to provide floor information in elevators.} 
\label{fig:applications}
\end{figure}

In this paper, we investigated the possibility of conveying Braille characters using ultrasonic haptics and evaluated three haptic stimulation methods.
With the small sample size, we were not able to identify a clear difference between the methods, but we still see value in reporting the scores and the quantitative feedback, as well as the finding that haptic information can be conveyed with all three. 
The problem needs to be revisited with a larger sample size, to be able to draw clear conclusions about significant differences in the accuracy between the methods. 
Taking into account the presentation times and the expert feedback, the Point-by-Point and Row-by-Row methods could be beneficial in the initial Braille learning phase, whereas proficient users could potentially prefer the Constant method.
An interesting finding is that all participants reported no difficulties in transferring their Braille reading skills to the mid-air haptic interface, after only four training trails for each method. 
The participants had an overwhelmingly positive reaction to the system and provided a list of scenarios and concrete tasks in their everyday life that could potentially be facilitated by such device.
A selection of the possible applications is illustrated in Figure~\ref{fig:applications}. Note that further testing and development of the HaptiRead interface is required to achieve them.

\subsection{Limitations}
This first validation study was conducted using a small subset of Braille characters, limited to four cells, individually presented, to ensure internal validity and experimental control.
Further studies are required to validate the findings using a full 6-cell layout, as well as presenting the information in context (e.g. words and sentences).
Due to these limitations, it is difficult to compare the obtained results to the prior work.
Testing with 6-cell characters, might result in lower accuracy rates, they could, however, potentially be compensated by longer training sessions. 
Rendering 6 or even 8-cell Braille characters could potentially be facilitated in the future, by manifacturing mid-air haptic displays with smaller transducers and thus better spatial resolution.
In our extensive testing process with domain experts and users, we did not come across any major challenge or criticism that would pose a doubt that with sufficient testing and development, the HaptiRead system would not work for more complex information.

\section{Conclusion}
In this paper, we evaluate the possibility of using ultrasonic mid-air haptic technology to convey Braille. 
The obtained results hold importance for the field of Human Computer Interaction, because they provide the first empirical validation of employing mid-air haptics for developing interfaces for blind people.
We conduct performance and system usability tests and evaluate three different methods for generating the haptic stimulation.
Our results show that it is possible to convey Braille as touchless haptic stimulation in mid-air with all of the proposed methods.
The participants responded favorably to the concept, however, further testing and development is needed.
We hope that our study will spark research into using mid-air haptics to potentially make the everyday multisensory experience of visually impaired and blind people richer.

\section{Acknowledgements}
This research has received funding from the European Union's Horizon 2020 research and innovation programme under grant agreement \#737087 (Levitate).
\balance{}

\bibliographystyle{SIGCHI-Reference-Format}
\bibliography{sample}

\end{document}